\documentclass{article}%
\usepackage{amsfonts}
\usepackage{amsmath}
\usepackage{amssymb}
\usepackage{graphicx}%
\begin{document}

\title{Transmission time and resonant tunneling through barriers\\using localized quantum density soliton waves}
\author{Babur M. Mirza\\Department of Mathematics, \\Quaid-i-Azam University, Islamabad. 45320. PK.}
\maketitle

\begin{abstract}
In this paper, the interaction and transmission time of quantum density
solitons waves passing through finite barrier potentials is investigated.
Using the conservation of energy and of quantum density, it is first
demonstrated that the soliton waves possess the important particle-like
properties, including localization by a finite de Broglie wavelength and
constant uniform motion in free space. The passage of quantum density solitons
through barriers of finite energies is then shown to lead to the phenomena of
resonant tunneling and, in Josephson-like configurations, to the quantization
of magnetic flux. A precise general measure for barrier tunneling time is
derived which is found to give a new interpretation of the quantum
indeterminacy principles. 



\end{abstract}

\section{Introduction}

Particles in quantum theory are represented by Gaussian wave packets solutions
of the linear Schr\"{o}dinger equation. However it is well known that Gaussian
wave packets, in the absence of an external force, exhibit dispersion over
time. This lack of localizability is due to the fact that a Gaussian wave
packet is formed by a linear superposition of stationary states that decay
exponentially with time, hence collectively cause the wave packet dispersion
[1]. The only other known free wave packet solution of the linear
Schr\"{o}dinger equation is the Airy-Berry wave packet [2], which although
possesses the non-dispersion feature of quantum particles, however exhibits
acceleration in the absence of any external force.

It is shown here that non-spreading and other features of particles can be
described by a nonlinear solitary wave phenomenon using quantum potential
formulation of basic quantum theory. We show here that these soliton waves not
only remain localized in free space, but also possess a finite wavelength
equal to the de Broglie wavelength, and move with a constant uniform speed in
free space, hence preserve typical particle properties. We demonstrate the
physical validity of these soliton waves by applying them to a number of cases
that have been lab tested. In particular we derive a precise new measure of
particle tunneling time through a finite barrier, where despite many
alternative definitions a general measure of tunneling time is still lacking
(see for instance Refs. [3], and for further references on this topic).

The paper is organized as follows. First the coupled energy and continuity
equations, involving the quantum potential term, are solved for the localized
soliton waves in section 2. The solution gives the quantum density function as
the localized soliton wave, with wavelength equal to the de Broglie
wavelength, which moves at a constant uniform speed. We then apply, in section
3 and 4, the resulting soliton solution to study the case of a free particle
interacting with a finite potential well, where now the particle is
represented by the quantum density soliton. The correct tunneling condition,
as well as the flux quantization condition, are shown to follow simply from
the soliton wave behavior of particles. In section 5, a new measure for the
tunneling time is derived which can be regarded as a test of the proposed
soliton behavior. Finally, it is shown that the indeterminacy principles is a
direct consequence of the soliton phenomenon, with the natural assumption that
the transversal time must remain non-negative in the tunneling process. In the
following we shall keep to the one space dimension only, whose extension to
two or three dimensions is rather obvious.

\section{Quantum Density Solitons}

Quantum description of a system is given by the continuity equation for the
quantum density function $\rho(x,t)$:
\begin{equation}
\frac{\partial\rho(x,t)}{\partial t}+\frac{\partial\left[  \rho
(x,t)u(x,t)\right]  }{\partial x}=0, \label{1}%
\end{equation}
and the energy equation: $E=\frac{1}{2}mv^{2}+Q(x,t)+V(x)$, involving the
quantum potential $Q(x,t)=-(\hbar^{2}/2m)(\partial^{2}\sqrt{\rho
(x,t)}/\partial x^{2})/\sqrt{\rho(x,t)}$, and the classical potential function
$V(x)$. The energy condition corresponds to the dynamical equation:%
\begin{equation}
m\left[  \frac{\partial u(x,t)}{\partial t}+u(x,t)\frac{\partial
u(x,t)}{\partial x}\right]  =-\frac{\partial Q(x,t)}{\partial x}%
-\frac{\partial V(x)}{\partial x}, \label{2}%
\end{equation}
where $u(x,t)$ is the particle/wave speed [4-7]. Here the term $\partial
V(x)/\partial x$ is the classical force, which vanishes for the free particle
case, thus $V$ is a constant.

Introducing the traveling wave variable $\xi=x-ct$, such that $\rho
(x,t)=\rho(\xi)$, and $u(x,t)=u(\xi)$ and re-writing $Q(x,t)$, as the quantum
potential divided by the mass $m$, we obtain:%
\begin{equation}
\rho(\xi)u^{\prime}(\xi)+\rho^{\prime}(\xi)u(\xi)=c\rho^{\prime}(\xi),
\label{3}%
\end{equation}%
\begin{equation}
u(\xi)u^{\prime}(\xi)+Q\prime(\xi)=cu^{\prime}(\xi), \label{4}%
\end{equation}
where the prime denotes differentiation with respect to the travelling wave
variable $\xi$.

We now solve equations (3) and (4) for the localized soliton waves. Although a
large class of soliton waves have been discovered, in a variety of nonlinear
problems, localized (or compact) soliton waves correspond to nonspreading
waves of finite wavelength, free from spatial extensions, such as exponential
tails or wings [8]. This is due to the restriction on the domain of periodic
functions by a compact support.

For the QHD equations, we assume that the quantum density function $\rho$
represents the particle-like localization, therefore, we take the ansatz for
the soliton solution:
\begin{align}
\rho(\xi)  &  =\rho_{0}\cos^{\beta}(\mu\xi),\text{ \ \ }\mid\mu\xi\mid
<\pi/2,\nonumber\\
&  =0,\text{ \ elsewhere.} \label{5}%
\end{align}
Here the inequality $\mid\mu\xi\mid<\pi/2$ represents the compact support for
the cosine function, which localizes the quantum density. Putting $\rho(\xi)$
from equation (5) in equations (3) and (4) yields $u(\xi)=c$ and $\beta=0,2$
and $\mu=\sqrt{A}$ where the constant of integration $A$ is to be identified,
using equation (4), as the initial form of the quantum potential
$Q(0,0)=Q_{0}$.

This implies that the quantum hydrodynamic equations allow a constant quantum
density (for $\beta=0$), as in the usual linear quantum theory. However there
is another possible solution corresponding to $\beta=2$, which yields a
localized soliton density wave, traveling with a constant uniform speed $c$.
Thus in the cosine representation of the solitary wave, the localized soliton
solution is given by:%
\begin{align}
\rho(\xi)  &  =\rho_{0}\cos^{2}\frac{\sqrt{2mQ_{0}}}{\hbar}\xi,\text{
\ \ }\mid\mu\xi\mid<\pi/2,\nonumber\\
&  =0,\text{ \ elsewhere.} \label{6}%
\end{align}
It is easily verified that the quantum density function (6), with the wave
speed $u(\xi)=c$, is a solution to the QHD equations (3) and (4), and
equivalently equations (1) and (2).

Indeed equation (2) implies that $Q_{0}=E-V$, and thus the wavelength of the
soliton must be equal to the de Broglie wavelength $h/\sqrt{2m(E-V)}$. The
localized soliton has a constant amplitude, equal to $\rho_{0}$, which is
independent of the soliton wavelength. Also the soliton wave moves in free
space, along the $x$ direction, with a constant uniform speed $c$. Notice that
the quantum potential for the quantum density soliton comes out a constant,
therefore the quantum force corresponding to the quantum density soliton is
identically zero.

\section{Application to Resonant Tunneling}

We now apply the above obtained results to the case of resonant tunneling
through a barrier of finite height. It is demonstrated that without making any
extra assumptions, only the requirement that the quantum density remains
continuous throughout the process, especially at the barrier walls, it is
possible to deduce the correct resonant tunneling conditions. In what follows
we use the terms particle and quantum density soliton interchangeably.

Consider a freely moving quantum density soliton interacting with a barrier
potential of width $a$ and height $V_{0}$:%
\begin{align}
V(x)  &  =V_{0},\text{ }0<x<a,\nonumber\\
&  =0,\text{ \ \ otherwise}. \label{7}%
\end{align}
Refer as region II to the region $0<x<a$, where the potential has the constant
value $V_{0}$, and region I and III as free regions to the left and right side
of the barrier, respectively. Such a potential occurs, for instance, in the
Josephson junction with the superconductor-insulator-superconductor
configuration [9].

The quantum density soliton inside the insulating region has a momentum $\hbar
k_{2}=\sqrt{2m(E-V_{0})}$. Now two different cases may arise: either $E>V_{0}%
$, or $E<V_{0}$.

For $E>V_{0}$, the quantum potential in region II is $Q=\hbar^{2}k_{2}%
^{2}/2m>0$, therefore the quantum density in region II is a soliton wave given
by:
\begin{equation}
\rho_{II}(\xi)=\rho_{2}\cos^{2}(k_{2}\xi). \label{8}%
\end{equation}
In the left and right hand regions the density is given respectively by the
free quantum solitons:
\begin{equation}
\rho_{I}(\xi)=\rho_{1}\cos^{2}(k_{1}\xi), \label{9}%
\end{equation}%
\begin{equation}
\rho_{III}(\xi)=\rho_{3}\cos^{2}(k_{3}\xi), \label{10}%
\end{equation}
with the particle/soliton momenta $\hbar k_{1}$, and $\hbar k_{3}$
respectively. Notice that, if however $E<V_{0}$, the density function in
region II is $\rho_{II}(\xi)=\rho_{2}\cosh^{2}(k_{2}\xi)$, where $\hbar
k_{2}=\sqrt{2m(V_{0}-E)}$.

Now the density wave amplitude $\rho_{2}$, in region II, can be calculated
from the incident wave amplitude, and the barrier parameters, as follows.
First using the continuity condition, at $x=0$, we have $\rho_{I}=\rho_{II}$,
from which it follows that at the initial wall,
\begin{equation}
\rho_{2}=\rho_{1}, \label{11}%
\end{equation}
Similarly at the other end of the barrier at wall $x=a$, we obtain:%
\begin{equation}
\rho_{3}=\frac{\cos^{2}(k_{2}a)}{\cos^{2}(k_{1}a)}\rho_{2},\text{ } \label{12}%
\end{equation}
Putting the matching conditions (11) and (12), in equations (8) to (10), it
follows that the quantum density soliton wave of given energy, after
interacting with the potential barrier, identically recovers itself (that is
$\rho_{I}=\rho_{III}$, and $k_{1}=k_{3}$), only when $k_{1}=k_{2}\pm n\pi$ for
$n=0,1,2,...$. These are the conditions for resonant tunneling through the
barrier. Similarly when $E<V_{0}$, we obtain the same tunneling conditions.

\section{Flux Quantization}

The Josephson junction provides another important example where the above
considerations of quantum density solitons are of rather direct significance.
In the Josephson junction [9], the superconductor region is modeled by\ the
macroscopic wave function $\psi(x,t)=\sqrt{n_{e}}e^{i\theta(x,t)}$, where
$n_{e}$ is the number density of the superconducting electrons. In the quantum
potential formalism, this wave function corresponds to the quantum density
$\rho(x,t)=n_{e}(x,t)$.

Thus with an initially normalized number density $n_{e}$, in the
superconducting region III, it follows from equations (8) to (12) that the
transmitted current density for $E>V_{0}$ is:%
\begin{equation}
n_{e}(x,t)=\frac{\cos^{2}(\sqrt{2m(E-V_{0})}a/\hbar)}{\cos^{2}(\sqrt
{2mE}a/\hbar)}\cos^{2}\frac{\sqrt{2mE}}{\hbar}(x-ct), \label{13}%
\end{equation}
and similarly if $E<V_{0}$,%
\begin{equation}
n_{e}(x,t)=\frac{\cosh^{2}(\sqrt{2m(V_{0}-E)}a/\hbar)}{\cos^{2}(\sqrt
{2mE}a/\hbar)}\cos^{2}\frac{\sqrt{2mE}}{\hbar}(x-ct), \label{14}%
\end{equation}
Since energy is the time derivative of the phase function, we have
$E=-\hbar\partial\theta_{1}/\partial t$ and $\left(  E-V_{0}\right)
=-\hbar\partial\theta_{2}/\partial t$, whereas $V_{0}$ is the junction potential.

The above formulas imply that there will be a conduction current across the
junction without loss for $E>V_{0}$ if,%
\begin{equation}
\frac{\sqrt{2mE}a}{\hbar}=\frac{\sqrt{2m(E-V_{0})}a}{\hbar}\pm n\pi,
\label{15}%
\end{equation}
and for $E<V_{0}$ if,%
\begin{equation}
\cos^{2}(\frac{\sqrt{2mE}a}{\hbar})=\cosh^{2}(\frac{\sqrt{2m(V_{0}-E)}a}%
{\hbar}).\text{ } \label{16}%
\end{equation}

We now show that flux quantization follows from equation (15), or equivalently
from equation (16). Writing equation (15) in terms of momenta $p_{1}%
=\sqrt{2mE}$ and $p_{2}=\sqrt{2m(E-V_{0})}$ implies that $(p_{1}-p_{2})a=\pm
n\pi\hbar$ , where for a full loop $a=2\pi$. Using now $p_{i}=\hbar
\nabla\theta_{i}$, where the index $i$ takes on values $1$ and $2$, we have on
integrating along the path from initial point with momentum $p_{1}$ and final
point with momentum $p_{2}$:%
\begin{equation}
\int_{1}^{2}\mathbf{\nabla}\theta_{1}\cdot ds-\int_{1}^{2}\mathbf{\nabla
}\theta_{2}\cdot ds=\pm\frac{n}{2}\int_{1}^{2}ds. \label{17}%
\end{equation}
Then for a complete loop, starting from say point $1$, going to the point $2$
and then ending at the starting point $1$, we have%
\begin{equation}%
{\displaystyle\oint}
\mathbf{\nabla}\theta\cdot ds=\pm n\pi. \label{18}%
\end{equation}
The left hand side of equation (18) defines the flux $\Phi$ through the closed
loop multiplied by $q/\hbar$ . Equation (18) is thus identical to the flux
quantization condition:
\begin{equation}
\Phi=\pm\frac{n\pi\hbar}{q}, \label{19}%
\end{equation}
where the flux $\Phi$ defined in terms of the vector potential $\mathbf{A}$ is
given by $\Phi=%
{\displaystyle\oint}
\mathbf{A}\cdot ds$. Notice that in deducing condition (19), the usual factor
of $2$ does not appear, hence we need not re-define charge in terms of Cooper pairs.

\section{Tunneling Time and the Indeterminacy Principle}

Quantum tunneling time through barriers has been investigated in different
contexts [10-14]\textit{. }We now apply the soliton wave representation to
estimate the time of tunneling through a potential barrier.

Referring to the potential barrier considered in section 3 above, let $x_{1}$,
$x_{2}$, and $x_{3}$ be the particle/soliton positions at three arbitrary
points in region I, II, and III, respectively. \ Let the respective time
instants of soliton arrival at these positions be $t_{1}$, $t_{2}$, and
$t_{3}$. After interaction the soliton is free, therefore has the same quantum
density $\rho_{I}=\rho_{III}$, however the wave number, as well as the
position and the time coordinated are different. Thus we have in this case
$\rho_{I}(\xi_{1})=\rho_{III}(\xi_{3})$, or explicitly, $\rho_{1}\cos^{2}%
k_{1}\xi_{1}=\rho_{3}\cos^{2}k_{3}\xi_{3}$. Then employing the matching
conditions (11) and (12), it follows, that the time for particle/soliton to
arrive at point $x_{3}$ is given by:%
\begin{equation}
t_{3}=\frac{x_{3}}{c}-\frac{1}{k_{3}c}\cos^{-1}\left[  \frac{\cos(k_{1}%
a)}{\cos(k_{2}a)}\cos k_{1}\left(  x_{1}-ct_{1}\right)  \right]  , \label{20}%
\end{equation}
where $a$ is the barrier width. Thus if initially the particle/soliton was at
$x_{1}=0$, at time $t=0$, then the time it arrives, at the barrier wall
$x_{3}=a$, is:%
\begin{equation}
t_{3}=\frac{a}{c}-\frac{1}{k_{3}c}\cos^{-1}\left[  \frac{\cos(k_{1}a)}%
{\cos(k_{2}a)}\right]  . \label{21}%
\end{equation}
This is the total time a quantum density soliton takes to transverse a barrier
of length $a$. In this formula the first term corresponds to the classical
transit time, whereas the second term involves the effects of the barrier.
Since tunneling time is the time during which the soliton-barrier interaction
takes place, it must be the residual (in absolute measures) $\mid t_{3}%
-\frac{a}{c}\mid$, that is, the difference between the total transversal time
$t_{3}$, and the time $a/c$ it takes to travel the same distance with the same
given speed, when free. Therefore according to equation (21) the quantum
tunneling time, denoted by $\tau$, is given by%
\begin{equation}
\tau=\frac{1}{k_{3}c}\cos^{-1}\left[  \frac{\cos(k_{1}a)}{\cos(k_{2}%
a)}\right]  . \label{22}%
\end{equation}
In formula (22), two points are noticed:

(1) $\tau$ is an oscillatory function, thus it can delay as well as shorten
the tunneling time. Also, due to this oscillatory behavior, it can be zero
under appropriate conditions that are the same as for resonant tunneling, that
is $k_{1}=k_{2}\pm n\pi$ where $n=0,1,2,...$.

(2) The denominator in this formula corresponds to the frequency of the
outgoing free particle, that is $k_{3}c=\nu_{3}$. This has implications
regarding whether tunneling\ can be an instantaneous process. Excluding the
resonant tunneling case, this requires that $\tau=0$. For the free particle
speed $c$ is finite, this requirement implies $k_{3}$, thus the frequency for
free outgoing particle, is infinite, hence it must carry an infinite energy.
Thus particles/quantum density solitons cannot transverse a barrier of finite
length instantaneosly.

The measure of tunnelining time (21) consistently recovers the classical
transversal time.

\subsection{The Indeterminacy Principle}

A direct consequence of formula (20) is the indeterminacy principle, which is
now deduced from the condition that the total time $t_{3}$ in formula (20) (or
equivalently formula (21)) cannot be negative or zero for a barrier of finite,
non-zero width $a$. Thus we take $t_{3}\geqslant0$ in equation (20). This
gives the condition%
\begin{equation}
x_{3}\geq\frac{1}{k_{3}}\cos^{-1}\left[  \frac{\cos(k_{1}a)}{\cos(k_{2}a)}\cos
k_{1}\left(  x_{1}-ct_{1}\right)  \right]  \label{23}%
\end{equation}
for each value of $k_{1},k_{2},k_{3}$ and $a$, and for each choice of $x_{1}$,
and $t_{1}$. This implies that maximally $x_{3}\geqslant\pi/k_{3}$. Replacing
coordinate $x_{3}$ by $\Delta x$, that is the distance measured from the
reference point $x_{1}$, of the point $x_{3}$, we obtain $\Delta x\geqslant
\pi/\Delta k$, and if the quantum density soliton after interaction has
momentum $\hbar\Delta k=\Delta p$, it follows that:%
\begin{equation}
\Delta x\Delta p\geqslant h/2, \label{24}%
\end{equation}
where $h=2\pi\hbar$ is the Planck's constant.\ In circular measures (23) has
the standard form: $\Delta x\Delta p\geqslant\hbar/2$. Similar considerations
give the energy-time indeterminacy relation also.

\section{Conclusions}

The above analysis shows how particle-like phenomena at the quantum scale can
be described in terms of localized soliton waves of de Broglie wavelength.
These soliton waves do not form as a linear superposition of the plane wave
solutions (stationary states) of the Schr\"{o}dinger equation, hence cannot be
represented as a Fourier sum, but must be considered as a (non-identical)
consequence of the quantum potential based formalism (1) and (2). The validity
of the soliton wave solutions is established by the correct tunneling, and the
flux quantization conditions. Furthermore, we have derived a new measure of
the time for particle to tunnel through a barrier, which leads to the
indeterminacy principle as a consequance.

Finally, inequality (23) can now be interpreted as follows. The quantum
density soliton has wavelength $h/\Delta p$, whereas the distance covered by
the soliton is $\Delta x$. The factor $1/2$, on the right hand side of
expression (23) is due to the symmetry (in cosine function) of soliton
wavelength $\lambda$, about the mean (reference) position, whereas $\Delta x$
measures length only on one side of the mean position. Inequality (23) thus
states that: the length to be transversed must be greater than or equal to the
wavelength of the quantum density wave itself.


\begin{thebibliography}{99}                                                                                               %


\bibitem {[1]}D. Bohm, \textit{Quantum Theory. }(Dover, New York, 1989), pp.65-68.

\bibitem {[2]}V. M. Berry and N. L. Balazs, Am. J. Phys. \textbf{47}, 264 (1979).

\bibitem {[3]}P. C. W. Davies, arXiv:quant-ph/0403010 (2004).

\bibitem {[4]}V. E. Madelung, Z. Physik. \textbf{40}, 322 (1926).

\bibitem {[5]}L. de Broglie, \textit{Electrons et Photons, Report au Ve
Conseil Physique Solvay} (Gauthier-Villiars, Paris, 1930).

\bibitem {[6]}D. Bohm, Phys. Rev. \textbf{85}, 166 (1952),; also D. Bohm,
Phys. Rev. \textbf{85}, 180 (1952).

\bibitem {[7]}Mirza, B. M., J.Phys.: Conf. Ser. \textbf{490}, (2014), p. 012198.

\bibitem {[8]}P. Rosenau and J.M. Hyman, Phys. Rev. Lett. \textbf{70}, 564-567 (1993).

\bibitem {[9]}T. P. Orlando and K. P. Delin, \textit{Foundations of Applied
Superconductivity.} (Addison-Wesley, New York, 1991). Ch.8.

\bibitem {[10]}D. Shafir, et.al., Nature. \textbf{485}, 343 (2012). \ 

\bibitem {[11]}R. W. Boyd and D. J. Gauthier, Science. \textbf{326}, 1074 (2009).

\bibitem {[12]}P. Eckle, et. al., Science. \textbf{322}, 1325 (2008). \ 

\bibitem {[13]}M. Uiberacker, et.al., Nature. \textbf{446}, 627 (2007).

\bibitem {[14]}M. Hentschel, et.al., Nature. \textbf{414}, 509 (2001).
\end{thebibliography}
\end{document}